\documentclass[12pt,preprint,natbib,iop]{emulateapj} 
\usepackage{graphicx,epsfig,amsmath} 

\lefthead{van der Wel}
\righthead{Structure of star-forming galaxies at high redshift.}
\slugcomment{Version: \today}
\begin{document}

\newcommand{\gala}{{\tt GALAPAGOS}~}
\newcommand{\gf}{{\tt GALFIT}~}
\newcommand{\mh}{H_{\rm{F160W}}}
\newcommand{\kms}{\>{\rm km}\,{\rm s}^{-1}}
\newcommand{\reff}{R_{\rm{eff}}}
\newcommand{\msol}{M_{\odot}}
\newcommand{\msola}{10^{11}~M_{\odot}}
\newcommand{\msolb}{10^{10}~M_{\odot}}
\newcommand{\msolc}{10^{9}~M_{\odot}}

\title{Geometry of Star-Forming Galaxies from SDSS, 3D-HST and
  CANDELS} \footnote{Based on observations with the Hubble Space
  Telescope, obtained at the Space Telescope Science Institute, which
  is operated by AURA, Inc., under NASA contract NAS 5-26555.}

\author{A.~van der Wel\altaffilmark{1}}
\author{Yu-Yen Chang\altaffilmark{1}}
\author{E.~F.~Bell\altaffilmark{2}}
\author{B.~P.~Holden\altaffilmark{3}}
\author{H.~C.~Ferguson\altaffilmark{4}}
\author{M.~Giavalisco\altaffilmark{5}}
\author{H.-W.~Rix\altaffilmark{1}}
\author{R.~Skelton\altaffilmark{6}}
\author{K.~Whitaker\altaffilmark{7}}
\author{I.~Momcheva\altaffilmark{8}}
\author{G.~Brammer\altaffilmark{4}}
\author{S.~A.~Kassin\altaffilmark{4}}
\author{M.~Martig\altaffilmark{1}}
\author{A.~Dekel\altaffilmark{9}}
\author{D.~Ceverino\altaffilmark{10}}
\author{D.~C.~Koo\altaffilmark{3}}
\author{M.~Mozena\altaffilmark{3}}
\author{P.~G.~van Dokkum\altaffilmark{8}}
\author{M.~Franx\altaffilmark{11}}
\author{S.~M.~Faber\altaffilmark{3}}
\author{J.~Primack\altaffilmark{12}}

\altaffiltext{1}{Max-Planck Institut f\"ur Astronomie, K\"onigstuhl
  17, D-69117, Heidelberg, Germany; e-mail:vdwel@mpia.de}

\altaffiltext{2}{Department of Astronomy, University of Michigan, 500
  Church Street, Ann Arbor, MI 48109, USA}

\altaffiltext{3}{UCO/Lick Observatory, Dept. of Astronomy and
  Astrophysics, University of California, Santa Cruz, CA 95064 USA}

\altaffiltext{4}{Space Telescope Science Institute, 3700 San Martin
  Drive, Baltimore, MD 21218, USA}

\altaffiltext{5}{Astronomy Department, University of Massachusetts, Amherst, MA, 01003, USA}

\altaffiltext{6}{South African Astronomical Observatory, PO Box 9,
  Observatory, 7935, South Africa}

\altaffiltext{7}{Astrophysics Science Division, Goddard Space Center,
  Greenbelt, MD 20771, USA}

\altaffiltext{8}{Department of Astronomy, Yale University, New Haven,
  CT 06511, USA}

\altaffiltext{9}{Center for Astrophysics and Planetary Science, Racah
  Institute of Physics, The Hebrew University, Jerusalem 91904,
  Israel}

\altaffiltext{10}{Department of Theoretical Physics, Universidad Autonoma
  de Madrid, E-28049 Madrid, Spain}

\altaffiltext{11}{Leiden Observatory, Leiden University, P.O.Box 9513,
  NL-2300 AA Leiden, The Netherlands}

\altaffiltext{12}{Department of Physics, University of California at
  SantaCruz, Santa Cruz, CA 95064, USA}

\begin{abstract}
  We determine the intrinsic, 3-dimensional shape distribution of
  star-forming galaxies at $0<z<2.5$, as inferred from their observed
  projected axis ratios.  In the present-day universe star-forming
  galaxies of all masses $10^9 - \msola$ are predominantly thin,
  nearly oblate disks, in line with previous studies.  We now extend
  this to higher redshifts, and find that among massive galaxies
  ($M_*>\msolb$) disks are the most common geometric shape at all
  $z\lesssim 2$.  Lower-mass galaxies at $z>1$ possess a broad range
  of geometric shapes: the fraction of elongated (prolate) galaxies
  increases toward higher redshifts and lower masses. Galaxies with
  stellar mass $\msolc$ ($\msolb$) are a mix of roughly equal numbers
  of elongated and disk galaxies at $z\sim 1$ ($z\sim 2$).  This
  suggests that galaxies in this mass range do not yet have disks that
  are sustained over many orbital periods, implying that galaxies with
  present-day stellar mass comparable to that of the Milky Way
  typically first formed such sustained stellar disks at redshift
  $z\sim1.5-2$.  Combined with constraints on the evolution of the
  star formation rate density and the distribution of star formation
  over galaxies with different masses, our findings imply that,
  averaged over cosmic time, the majority of stars formed in disks.
\end{abstract}

\section{Introduction}\label{sec:intro}
The shape of the stellar body of a galaxy reflects its formation
process.  Reconstructing the intrinsic, three-dimensional shapes of
spiral galaxies from their shapes projected on the sky has a long
tradition, and proved to be an exquisitely accurate and precise
approach, especially once sample size increased
\citep[e.g.,][]{sandage70, lambas92, ryden04, vincent05, padilla08}.
These results provided us with the general notion that the stellar
bodies of present-day star-forming galaxies over a wide range in
luminosity can be described as thin, nearly oblate (therefore,
disk-like) systems with an intrinsic short-to-long axis ratio of
$\sim0.25$.  Such global shapes encompass all galactic components,
including bars and bulges. The disk component is generally thinner
\citep[$0.1-0.2$, e.g.,][]{kregel02}.

Analogous information about the progenitors of today's galaxies is
scarcer.  Among faint, blue galaxies in deep Hubble Space Telescope
imaging, \citet{cowie95} found a substantial population of elongated
`chain' galaxies, but several authors argued that chain galaxies are
edge-on disk galaxies \citep[e.g.,][]{dalcanton96, elmegreen04a,
  elmegreen04b}.  However, \citet{ravindranath06} demonstrated that
the ellipticity distribution of a large sample of $z=2-4$ Lyman Break
Galaxies is inconsistent with randomly oriented disk galaxies, lending
credence to the interpretation that a class of intrinsically elongated
(or, prolate) objects in fact exists at high redshift.  By modeling
ellipticity distributions, \citet{yuma11} and \citet{law12} concluded
that the intrinsic shapes of $z>1.5$ star-forming galaxies are
strongly triaxial.

On the other hand, regular rotation is commonly seen amongst $z\sim
1-2$ samples \citep{forster06, kassin07, law09, forster09,
  wisnioski11, gnerucci11, newman13}, and the evidence for the
existence of gaseous disks is ample among massive systems
\citep{genzel06, wright07, lottie08, stark08, epinat09}.
One possible explanation for the seeming discrepancy between the
geometric and kinematic shape inferences is a dependence of structure
on galaxy mass.  Indeed, for lower-mass galaxies ($\lesssim \msolb$)
the evidence for rotation is less convincing
\citep[e.g.,][]{forster06, law07} and in rare cases rotation is
convincingly ruled out \citep[e.g.,][]{lowenthal09}.  The prevailing
view is that the gas --and hence presumably the stars that form from
it -- in those galaxies is supported by random motions rather than
ordered rotation.  However, the kinematic measurements for low-mass
galaxies probe only a small number of spatial resolution elements --
signs of rotation may be smeared out \citep{jones10} -- and the
observed motions may have a non-gravitational origin such as feedback.

Here we aim to provide the first description of the geometric shape
distribution of $z>1$ star-forming galaxies and its dependence on
galaxy mass.  We examine the projected axis ratio distributions
($p(q)$) of large samples of star-forming galaxies out to $z=2.5$
drawn from the CANDELS \citep{grogin11, koekemoer11} and 3D-HST
\citep{brammer12, skelton14} surveys.  A low-redshift comparison
sample is drawn from the Sloan Digital Sky Survey (SDSS).  The
methodology developed by \citet{holden12} and \citet{chang13b} will be
used to convert $p(q)$ into 3-dimensional shape distributions of
star-forming galaxies and its evolution from $z=2.5$ to the present
day.

\begin{figure*}[t]
\epsscale{1.2} 
\plotone{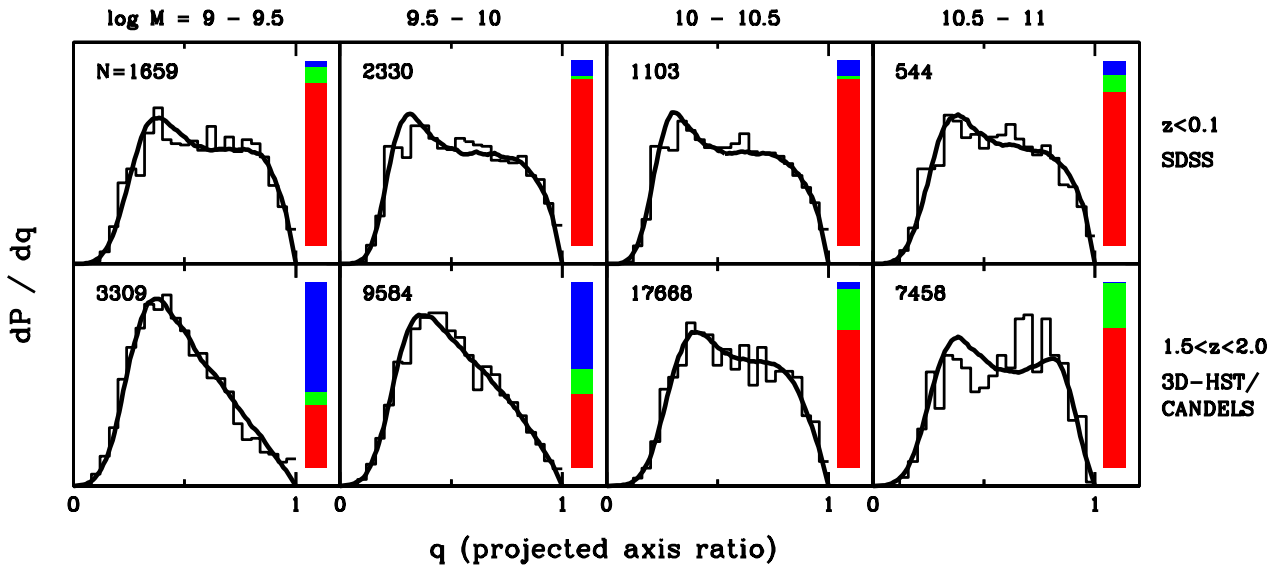}
\caption{Projected axis ratio distributions $p(q)$ of star-forming
  galaxies in four mass bins and two redshift bins ($z<0.1$ at the
  top; $1.5<z<2.0$ at the bottom).  Histograms represent the observed
  distributions (each panel contains 500 or more galaxies); continuous
  lines are best-fitting models: these are the probability
  distributions of triaxial populations of objects seen at random
  viewing angles., where the triaxiality and ellipticity are tuned to
  best reproduce the observed distributions.  The colored bars
  illustrate how the model populations are distributed over three
  different 3D shapes defined in Figure \ref{class}: \emph{disky} in
  red; \emph{spheroidal} in green; \emph{elongated} in blue.  The
  pronounced variation among the projected axis ratio distributions
  illustrates that the changes in the geometric fractions are highly
  significant.}
\label{hist}
\end{figure*}

\begin{figure}[t]
\epsscale{1.2} 
\plotone{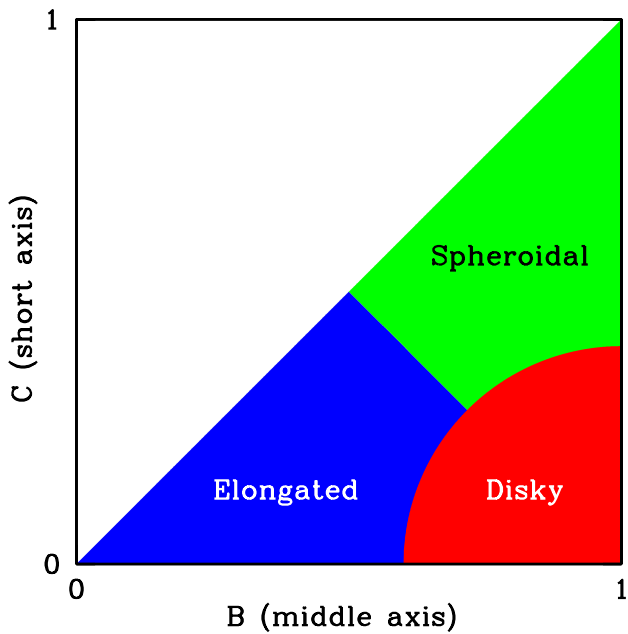}
\caption{To facilitate a better intuitive understanding of the model
  shape parameters (triaxiality and ellipticity) we distinguish three
  crudely defined 3-dimensional shapes of objects.  Objects with three
  similarly long axes are defined as \emph{spheroidal}; objects with
  two similarly long and one short axis are defined as \emph{disky};
  objects with one long axis and two similarly short axes are defined
  as \emph{elongated}.  A model population -- generated to reproduce
  an observed axis-ratio distribution -- should be thought of as a
  cloud of points in the parameter space shown in this figure,
  distributed as prescribed by the best-fitting values of $T$,
  $\sigma_T$, $E$, and $\sigma_E$ (see text for details).  Each of the
  three regions will contain a given fraction of those points, that
  is, a fraction of the population. }
\label{class}
\end{figure}

\section{Data}\label{sec:data}

We construct volume-limited samples of star-forming galaxies over a
large range in stellar mass ($10^9 - \msola$) and redshift ($0<z<2.5$)
with $q$ measured at an approximately fixed rest-frame wavelength of
$4600\rm{\AA}$.

\subsection{CANDELS and 3D-HST}
\citet{skelton14} provide WFC3/F125W+F140W+F160W-selected,
multi-wavelength catalogs for the CANDELS fields, as well as
redshifts, stellar masses and rest-frame colors using the 3D-HST WFC3
grism spectroscopy in addition to the photometry.  36,653 star-forming
galaxies with stellar masses $M_*>\msolc$ and up to redshift $z=2.5$
are selected based on their rest-frame $U-V$ and $V-J$ colors as
described by \citet{vanderwel14}, 35,832 of which have $q$
measurements.  The typical accuracy and precision is better than 10\%
\citep{vanderwel12}.  For the $2<z<2.5$ galaxies we use the
F160W-based values, for the $z<2$ galaxies we use the F125W-based
values, such that all $z>1$ galaxies have their shapes measured at a
rest-frame wavelength as close as possible to $4600\rm{\AA}$ (and
always in the range $4300<\lambda/\rm{\AA} <6200$).  This avoids the
effects due the shape variations with wavelength seen in local
galaxies \citep{dalcanton02}.

Below $z=1$ our F125W shape measurements probe longer wavelengths.  We
compared the F125W-based shapes with HST/ACS F814W-based shapes for
1,365 galaxies \citep[see][]{vanderwel14}.  The median F125W-based
axis ratio is 0.014 larger than the median F814W-based shape, with a
scatter of 0.06.  This is consistent with the measurement errors.  We
conclude that using F125W axis ratios at $z<1$ does not affect our
results.

\subsection{SDSS: $0.04<z<0.08$}

SDSS photometry-based stellar masses from \citet{brinchmann04} are
used to select 36,369 star-forming galaxies with stellar masses
$M_*>\msolc$ and in the (spectroscopic) redshift range $0.04<z<0.08$.
The distinction between star-forming and passive galaxies is described
by \citet{holden12} and is based on the rest-frame $u-r$ and $r-z$
colors, analogous to the use of $U-V$ and $V-J$ colors at higher
redshifts.  For the SDSS sample we use the $q$ estimates from fitting
the exponential surface brightness model to the $g$-band imaging as
part of the DR7 photometric pipeline \citep{abazajian09}.  These
measurements have been verified by \citet{holden12}, who showed that
systematic offsets and scatter with respect to our \gf-based
measurements are negligible.

\section{Reconstruction: from Projected to Intrinsic
  Shapes}\label{sec:model}

The very pronounced change of the projected shape distribution with
redshift (Figure \ref{hist}) immediately reveals that galaxy structure
evolves with cosmic time.  Especially at low stellar masses we see
that a larger fraction of galaxies have flat projected shapes than at
the present day.  This observation underpins the analysis presented in
the remainder of the Letter.  Here we provide a brief description of
the methodology to infer the intrinsic, 3-dimensional shapes of
galaxies, outlined in detail by \citet{chang13b}.

We adopt the ellipsoid as the general geometric form to describe the
shapes of galaxies.  It has three, generally different, axis lengths
($A \ge B \ge C$), commonly used to define ellipticity ($1-C$) and
triaxiality ($(1-B^2)/(1-C^2)$).  In order to facilitate an intuitive
understanding of our results we define three broad geometric types,
shown in Figure \ref{class}: \emph{disky} ($A \sim B > C$),
\emph{elongated} ($A > B \sim C$), and \emph{spheroidal} ($A \sim B
\sim C$).

The goal is to find a model population of triaxial ellipsoids that,
when seen under random viewing angles, has the same $p(q)$ as an
observed galaxy sample.  Our model population has Gaussian
distributions of the ellipticity (with mean $E$ and standard deviation
$\sigma_E$) and triaxiality (with mean $T$ and standard deviation
$\sigma_T$).  Such a model population has a known $p(q)$ which we
adjust to include the effect of random uncertainties in the axis ratio
measurements -- these are asymmetric for nearly round objects.  Then,
given that each observed value of $q$ corresponds to a known
probability, we calculate the total likelihood of the model by
multiplying the probabilities of each of the observed values.  We
search a grid of the four model parameters to find the maximal total
likelihood.

In Figure \ref{hist} we show observed axis ratio distributions
(histograms), and the probability distributions of the corresponding
best-fitting model populations (smooth lines).  The models generally
match the data very well.  Even in the worst case (bottom-right panel)
the model and data distributions are only marginally inconsistent, at
the $2\sigma$ level.
A triaxial model population with parameters $(E,\sigma_E,T,\sigma_T)$
corresponds to a cloud of points in Figure \ref{class} and, hence,
with certain fractions of the three geometric types.  The colored bars
in Figure \ref{hist} represent these fractions for the best-fitting
triaxial models.  This illustrates the connection between projected
shapes and intrinsic shapes: a broad $p(q)$ reflects a large fraction
of \emph{disky} objects, whereas a narrow distribution with a peak at
small $q$ is indicative of a large fraction of \emph{elongated}
objects.  A narrow distribution with a peak at large $q$ would
indicate a large fraction of \emph{spheroidal} objects.

\begin{figure*}[t]
\epsscale{1.2} 
\plotone{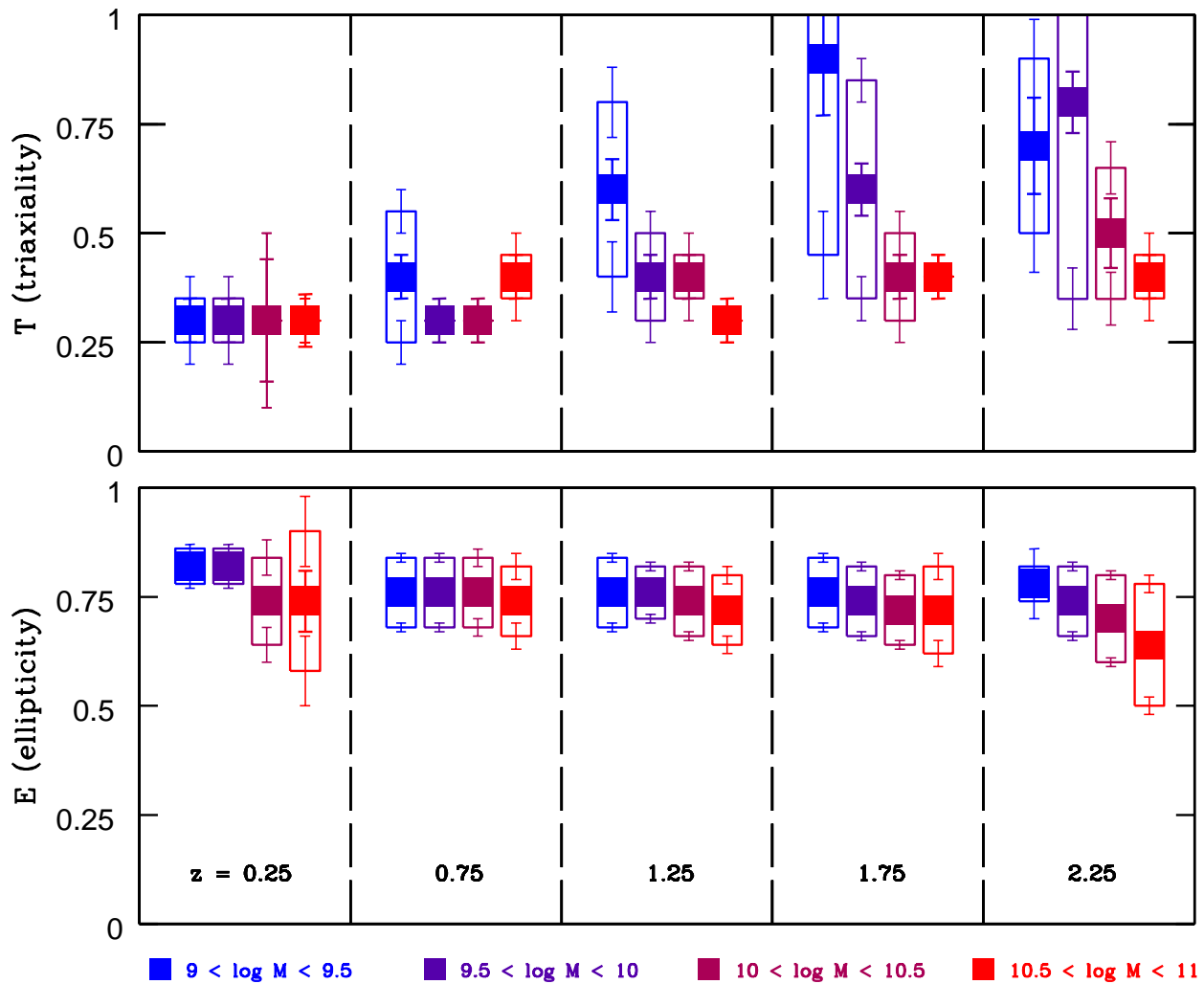}
\caption{Reconstructed intrinsic shape distributions of star-forming
  galaxies in our 3D-HST/CANDELS sample in four stellar mass bins and
  five redshift bins.  The model ellipticity and triaxiality
  distributions are assumed to be Gaussian, with the mean indicated by
  the filled squares, and the standard deviation indicated by the open
  vertical bars. The $1\sigma$ uncertainties on the mean and scatter
  are indicated by the error bars.  Essentially all present-day
  galaxies have large ellipticities, and small triaxialities -- they
  are almost all fairly thin disks.  Toward higher redshifts low-mass
  galaxies become progressively more triaxial.  High-mass galaxies
  always have rather low triaxialities, but they become thicker at
  $z\sim 2$.}
\label{res}
\end{figure*}

\begin{figure*}[t]
\epsscale{1.2} 
\plotone{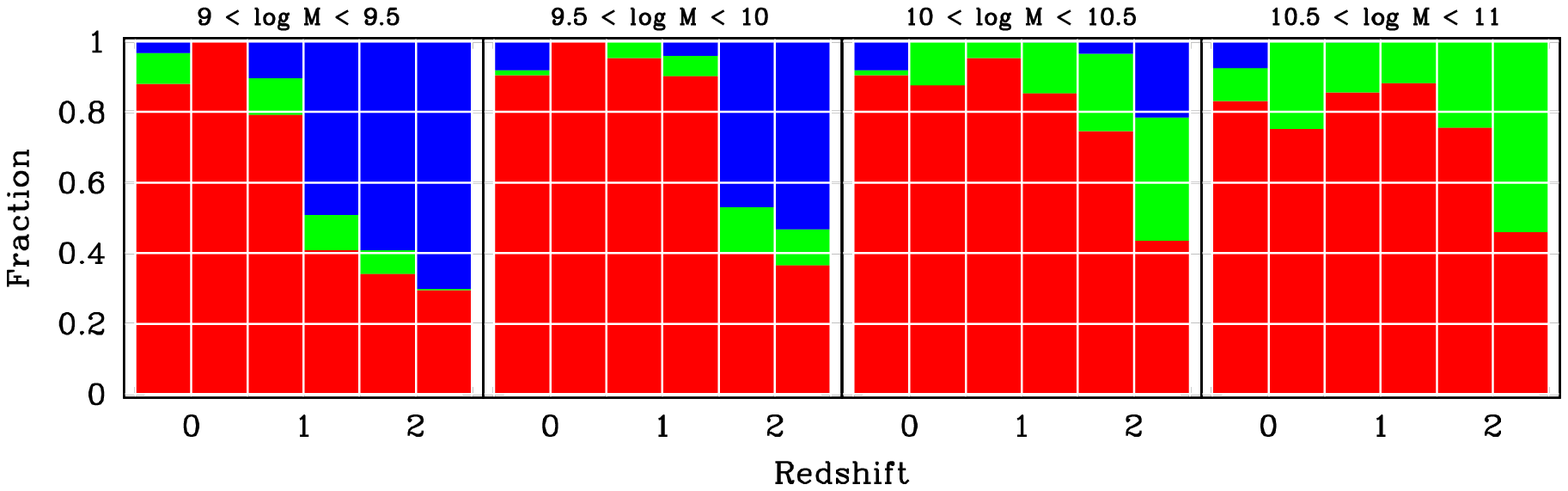}
\caption{Color bars indicate the fraction of the different types of
  shape defined in Figure \ref{class} as a function of redshift and
  stellar mass. The negative redshift bins represent the SDSS results
  for $z<0.1$; the other bins are from 3D-HST/CANDELS.}
\label{frac}
\end{figure*}

In Figure \ref{res} we provide the modeling results for the full
redshift and mass range probed here: for each stellar mass bin we show
the redshift evolution of the four model parameters, including the
uncertainties obtained by bootstrapping the samples.  Finally, in
Figure \ref{frac} we show the full set of results in the form of the
color coding defined in Figure \ref{class}.

\section{Evolution of Intrinsic Shape Distributions}

The small values of $T$ and the large values of $E$ for present-day
star-forming galaxies (Figure \ref{res}) imply that the vast majority
are thin and nearly oblate.  Indeed, according to our classification
shown in Figure \ref{hist} between 80\% and 100\% are \textit{disky},
as is generally known and was demonstrated before on the basis of
similar axis-ratio distribution analyses by \citet{vincent05} and
\citet{padilla08}.  Importantly, the intrinsic shape distribution of
star-forming galaxies does not change over a large range in stellar
mass ($10^9 - \msola$).

Toward higher redshifts star-forming galaxies become gradually less
disk-like (Figures \ref{hist}, \ref{res} and \ref{frac}).  This effect
is most pronounced for low-mass galaxies.  Already in the $0.5<z<1.0$
redshift bin in Figure \ref{res} we see evolution, mostly in the
scatter in triaxiality ($\sigma_T$). That is, there is substantial
variety in intrinsic galaxy shape.  Beyond $z=1$, galaxies with
stellar mass $\msolc$ typically do not have a \textit{disky} geometry,
but are most often \textit{elongated} (Figure \ref{res}).  Galaxies
with mass $\msolb$ show similar behavior, but with evolution only
apparent at $z>1.5$.  This geometric evidence for mass-dependent
redshift evolution of galaxy structure is corroborated by the analysis
of kinematic properties of $z=0-1$ galaxies by \citet{kassin12}.

\textit{Disky} objects are the most common type ($\ge75\%$) among
galaxies with mass $>\msolb$ at all redshifts $z\lesssim 2$.  A
population of \textit{spheroidal} galaxies is increasingly prominent
among massive galaxies at $z>2$.  A visual inspection of such objects
reveals that at least a subset are mergers, but an in-depth
interpretation of this aspect we defer to another occasion.

It is interesting to note that ellipticity hardly depends on mass and
redshift (Figure \ref{res}).  That is, despite strong evolution in
geometry, the short-to-long axis ratio remains remarkably constant
with redshift, and changes little with galaxy mass.  A joint analysis
of galaxy size and shape is required to explore the possible
implications.

Note that our definition of geometric shape is unrelated to the common
distinction between disks and spheroids on the basis of their
concentration parameter or S\'ersic index.  As a result we distinguish
between the observation that most low-mass star-forming galaxies at
$z\sim 2$ have exponential surface brightness profiles
\citep[e.g.,][]{wuyts11} and our inference that these galaxies are
not, generally, shaped like disks in a geometric sense.  This
illustrates that an approximately exponential light profile can
correlate with the presence of a disk-like structure but cannot be
used as a definition of a disk.

\section{Discussion}

Star formation in the present-day universe mostly takes place in
$>\msolc$ galaxies and in non-starburst galaxies.  Since essentially
all such star-forming galaxies are \textit{disky} and star formation
in disk galaxies occurs mostly over the full extent of the stellar
disk, it follows immediately that essentially all current star
formation takes place in disks.  The analysis presented in this
\textit{Letter} allows us to generalize this conclusion to include
earlier epochs.

At least since $z\sim 2$ most star formation is accounted for by
$\gtrsim \msolb$ galaxies \citep[e.g.,][]{karim11}.  Figures \ref{res}
and \ref{frac} show that such galaxies have disk-like geometries over
the same redshift range.  Given that 90\% of stars in the universe
formed over that time span, it follows that the majority of all stars
in the universe formed in disk galaxies.  Combined with the evidence
that star formation is spatially extended, and not, for example,
concentrated in galaxy centers \citep[e.g.,][]{nelson12, wuyts12} this
implies that the vast majority of stars formed in disks.

Despite this universal dominance of disks, the elongatedness of many
low-mass galaxies at $z\gtrsim 1$ implies that the shape of a galaxy
generally differs from that of a disk at early stages in its
evolution.  According to our results, an elongated, low-mass galaxy at
$z\sim 1.5$ will evolve into a disk at later times, or, reversing the
argument, disk galaxies in the present-day universe do not initially
start out disks.\footnote{This evolutionary path is potentially
  interrupted by the removal of gas and cessation of star formation.}

As can be seen in Figure \ref{res}, the transition from
\textit{elongated} to \textit{disky} is gradual for the population.
This is not necessarily the case for individual galaxies.
Hydrodynamical simulations indicate that sustained disks form quite
suddenly, on a dynamical time scale, after an initial period
characterized by rapidly changing dynamical configurations
\citep[e.g.,][]{martig14}.  This turbulent formation phase may include
the subsequent formation and destruction of short-lived disks
\citep[e.g.,][]{ceverino13}, associated with rapid changes in
orientation and resulting in a hot stellar system of rather arbitrary
shape.

Our observation that at $z>1$ the low-mass galaxy population consists
of a mix of \textit{disky} and \textit{elongated} objects -- in this
picture, the latter represent the irregular phase without a sustained
disk -- can be interpreted as some fraction of the galaxies having
already transformed into a sustained disk.  The probability for this
transition is, then, a function of mass which may or may not depend on
redshift.  Given the various estimates of the stellar mass evolution
of Milky Way-mass galaxies as a function of redshift
\citep[e.g.,][]{vandokkum13, patel13}, we suggest that the Milky Way
may have first attained a sustained stellar disk at redshift
$z=1.5-2$.

\section{Caveats}

Our analysis rests on the assumption that stellar light traces the
mass distribution of a galaxy.  Potential spoilers include obscuration
by dust, dispersion in age among stars, and large gas fractions.

Dust has a viewing angle-dependent effect on the measured $q$.
Massive galaxies at all redshifts are dusty, and a large variety of
dust geometries could disturb axis ratio measurements, hiding the
disk-like structure of the population when traced by the axis ratio
distribution. Perhaps this plays a role at $z>2$ where we see an
increased fraction of round objects.  However, the reverse -- to
create a disk-like axis ratio distribution for a population of dusty
non-disks -- requires unlikely fine tuning.  We prefer the more
straightforward interpretation that massive, star-forming galaxies
truly are disks, at least up to $z=2$.  This is supported by the
observed correlation between axis ratio and color
\citep[e.g.,][]{patel12}, also seen in our sample: galaxies with
smaller $q$ are redder than those with larger $q$, as expected from a
population on inclined, dusty disks.

Dust is also unlikely to affect $p(q)$ of low-mass galaxies. At $z>1$
galaxies with stellar masses $\lesssim \msolb$ are generally very
blue.  For these young, presumably metal-poor galaxies dust is of
limited relevance to the shape measurements.  This also implies that
completeness of our sample is not affected by strong dust obscuration.

Age variations in the stellar population and large gas fractions both
potentially present challenges to our assumption that the rest-frame
optical light traces the underlying mass distribution.  Perhaps
the luminous regions are young, bright complexes embedded in disks
consisting of cold gas or fainter, older stellar populations.  We
cannot immediately discard this possibility as dynamical masses exceed
stellar masses by an average factor of $\sim 3$ in the stellar mass
range $10^8~\msol \lesssim M_* \lesssim \msolb$ galaxies at $z>1$
\citep[e.g.,][]{forster09, maseda13}.

It is implausible that this difference between stellar mass and
dynamical mass is entirely made up of undetected, older stars in a
disk-like configuration.  The different spatial distributions of the
young and old stars would lead to wavelength-dependent shapes, which
is not observed.  If such a population of older stars is present, it
must be spatially coincident with the young population, and not,
generally, in a disk.

We cannot exclude the existence of cold gas disks that are
$\sim$3$\times$ more massive than the (young) stellar population.
Hydrodynamical simulations show that low-mass, high-redshift systems
can produce elongated stellar bodies embedded in more extended,
turbulent gaseous bodies with ordered rotation
\citep[e.g.,][]{ceverino13}.  At the moment there is little
observational evidence for such extended gaseous disks.  For the mass
range $10^{9.5}~\msol \lesssim M_* \lesssim \msolb$ gas masses in
excess of the stellar mass have been inferred based on the
star-formation rate and the inverse Kennicutt-Schmidt relation
\citep[e.g.,][]{forster09}, but this inversion relies on the
assumption of a disk-like geometry, weakening the argument.
Furthermore, even if these cold gas mass estimates are correct it is
not clear that the gas should be organized in a disk.  Generally, gas
ionized by star formation and cold gas share global kinematic traits,
and in these cases the ionized gas does not generally show rotation.
Deep ALMA observations will settle this issue, and for now we will
leave this as the main caveat in our analysis.

\section{Summary and Conclusions}

We have analyzed the projected axis ratio distributions, $p(q)$,
measured at rest-frame optical wavelenghts, of stellar mass-selected
samples of star-forming galaxies in the redshift range $0<z<2.5$ drawn
from SDSS and 3D-HST+CANDELS.  The intrinsic, 3-dimensional geometric
shape distribution is reconstructed under the assumption that the
population consists of triaxial objects view under random viewing
angles.

In the present-day universe star-forming galaxies of all masses are
predominantly oblate and flat, that is, they are disks.  Massive
galaxies ($M_*>\msolb$) typically have this shape at all redshifts
$0<z\lesssim 2$.  Given the dominance of $10^{10}-\msola$ galaxies in
terms of their contribution to the cosmic stellar mass budget and the
star formation rate density it follows that, averaged over all cosmic
epochs, the majority of all stars formed in disks.

Lower-mass galaxies have shapes at $z>1$ that differ significantly
from those of thin, oblate disks.  For galaxies with stellar mass
$\msolc$ ($\msolb$) there exists a mix of roughly equal numbers of
elongated and disk galaxies at $z\sim 1$ ($z\sim 2$).  At $z>1$ the
$\msolc$ galaxies are predominantly elongated.  Our findings imply
that low-mass galaxies at high redshift had not yet formed a regularly
rotating, sustained disk.  Given a range of plausible mass growth rate
of Milky Way-mass galaxies we infer the disk formation phase for such
galaxies at $z=1.5-2$.

\bibliographystyle{apj}

\end{document}